\documentclass[a4paper,10pt,twoside]{cpc-hepnp}

\usepackage{multicol}
\usepackage{graphicx}
\usepackage{booktabs}
\usepackage{amssymb,bm,mathrsfs,bbm,amscd}
\usepackage{verbatim}
\usepackage[tbtags]{amsmath}
\usepackage{lastpage}
\usepackage{CJK}

\begin{document}

\fancyhead[c]{\small Yu.G. Ignat'ev, A.A. Agathonov, I.A. Kokh} 
\fancyfoot[C]{\small 010201-\thepage}

\footnotetext[0]{Received 14 August 2018}

\title{The Peculiarities of the Cosmological Models Based on Non-Linear Classical and Phantom Fields with Minimal Interaction. I. The Cosmological Model Based on Scalar Singlet.\thanks{The work is performed according to the Russian Government Program of Competitive Growth of Kazan Federal University}}

\author{%
      Yurii Ignat'ev$^{1)},$\email{ignatev-yurii@mail.ru}%
\quad Alexander Agathonov$^{1,2)}$\email{a.a.agathonov@gmail.com}%
\quad and Irina Kokh$^{2)}$\email{irina\_kokh@rambler.ru}
}
\maketitle

\address{%
$^1$ Institute of Physics, Kazan Federal University,$^2$ Lobachevsky Institute of Mathematics and Mechanics, Kazan Federal University,  Kremleovskay str. 18, Kazan,  420008,  Russia\\
}

\begin{abstract}
A detailed comparative qualitative analysis and numerical simulation of evolution of the cosmological models based on classical and phantom scalar fields with self-action was performed. The phase portraits of the dynamic systems of classical and phantom fields and their projections to the Poincare sphere were constructed. It was shown that the phase trajectories of the corresponding dynamic systems can be split by bifurcation trajectories into 2,4 or 6 different dynamic streams corresponding to different pairwise symmetric histories of the Universe depending on the parameters of the scalar field's model. The phase space of such systems becomes multiply connected, the ranges of negative total effective energy unavailable for motion, getting appear there. In the case when attracting centers are situated inside these ranges, the phase trajectories of the classical scalar field in the infinite future tend to limit cycles, winding onto the boundaries of these ranges. The phase trajectories of the scalar field, in turn, get away from the boundaries of ranges with null effective energy and in the infinite future are wound onto one of the symmetrical focuses (centers). Thus, the situations when the Universe, in case of classical scalar field, begins its history with the inflation and ends it up in the Euclidian future, or, in the case of phantom scalar field, in opposite, has the Euclidian start and proceeds to inflation mode after anomalous burst of the acceleration, both become possible. The potentials of scalar fields on the surface of null curvature are distinct from zero and thereby define a certain vacuum state.
\end{abstract}



\footnotetext[0]{\hspace*{-3mm}\raisebox{0.3ex}{$\scriptstyle\copyright$}2013
Chinese Physical Society and the Institute of High Energy Physics
of the Chinese Academy of Sciences and the Institute
of Modern Physics of the Chinese Academy of Sciences and IOP Publishing Ltd}%

\begin{multicols}{2}

\section{Introduction}

 Standard cosmological models (SCM), (see, e.g., \cite{Gorb_Rubak}), based on classical scalar field were investigated using methods of qualitative analysis of dynamic systems in papers \cite{Belinsky,Zhur_01,Zhur,Mex1,Mex2}. In Zhuravlev's paper \cite{Zhur_01} the methods of qualitative theory of dynamic systems were also used for the research of a 2-component cosmological model with minimal interaction (see also \cite{Zhur}, \cite{Mex2}). In the Author's paper \cite{Ignat_16_1_stfi} the qualitative and numerical analysis of the standard cosmological model based on the classical scalar field was over again performed, now reducing the problem to the research of the dynamic system on the 2-dimensional phase plane $\{\Phi,\dot{\Phi}\}$ . The microscopic oscillating character of the invariant cosmological acceleration at late stages of the expansion was shown. The results  were then generalized to cosmological models with $\lambda$ - term \cite{Ignat_16_2_stfi}, \cite{Ignat_17_1_GC}, and the Authors managed to prove the conservation of the oscillating character of the invariant cosmological acceleration at sufficiently small values of the cosmological term. Besides that, in the recent papers the possibility of macroscopic acceleration's achieving non-relativistic mode at late stages of the early Universe\footnote{i.e. at stages with scalar field's prevalence} was shown by means of averaging the cosmological acceleration over microscopic oscillations.  . The effective macroscopic equation of state was obtained in  \cite{Ignat_Sam}, \cite{Ignat_Sam_Ignat} by means of direct averaging of the scalar potential and the derivative of the classical field over microscopic oscillations; it was also shown that it tends to non-relativistic state. Besides that, the macroscopic average of the scalar potential's square fluctuations was calculated in these works and it was also shown that at late stages of the cosmological evolution the root-mean-square energy of the microscopic oscillations of the scalar field exceeds the energy of the macroscopic scalar field by many orders  of magnitude.
In the work \cite{Zhur_16} the research method mentioned in the cited articles was used: it was applied to the 2-component system ``scalar field + liquid'' with an arbitrary function $V(\phi)$\footnote{In particular, for the Higgs potential.}.

\par In the historical perspective, the matter with negative pressure (creation $C$-field) was first introduced in the cosmology in the papers of F.Hoyle to maintain the Universe's stationary state \cite{Hoyle1949}, \cite{Hoyle1964}. Technically, the phantom fields, apparently were introduced in the gravity as one of the possible models of the scalar field in 1983 in the work \cite{Ignat83_1}. In the cited paper, as well as in the more recent (see e.g.  \cite{Ignat_Kuz84}, \cite{Ignat_Mif06}) the phantom fields were classified as scalar fields with attraction of like-charged particles being noted for the multiplier $\epsilon=-1$ in the scalar field's energy-momentum tensor. Let us notice, that phantom fields with respect to the wormholes and the so-called black Universes were considered in the papers \cite{Bron1}, \cite{Bron2}. In the given paper, in correspondence with the conventional terminology, under the phantom fields we will imply the scalar fields with negative kinetic term in the energy-momentum tensor irrespective of the sign of the potential term. Herewith, the negative potential term in the energy - momentum tensor correspond to the phantom scalar fields with attraction of the like-charged particles, and the positive potential term correspond to the phantom scalar fields with the repulsion. In the first case the signs of the kinetic term and the potential term coincide, in the second case they are opposite which is equivalent to the massive term's sign change in the Klein - Gordon equation. The corresponding solutions for the single-isolated scalar charge take not the form of Yukawa's potential but the form of the solutions of the Lifshitz scalar perturbations for the spherical symmetry ($\sin kr/r$) \cite{Ignat_12_3_Iz}.

\par However, the introduction of the phantom fields in the structure of the quantum field theory comes across serious problems related either to the probabilistic interpretation of the quantum theory, or to the problem of the vacuum state's stability due to unlimitedness of the negative energy \cite{Cline}. The negative kinetic term and the violation of the isotropic energy condition imply that the energy is not limited from below on the classical level, therefore negative norms appear on the quantum level. The negative norms of the quantum states, in turn, generate negative probabilities which contradict the standard interpretation of the quantum theory of field \cite{Linde}, \cite{Saridakis}. The requirement of the theory's unitarity leads to instability at description of interaction of the quantum phantom fields with other quantum fields \cite{Sbisa}.
However,  it is noted in the paper \cite{Vernov} that the terms that lead to the instability, can be considered as corrections, which are only significant at small energies below the level of the physical cutoff. Such an approach allows considering the phantom field theories as certain effective, physically acceptable theories where it is assumed that the effective theory allows inclusion into a certain fundamental theory, e.g. field string theory, which aligns with the well-known concept of S.M. Carroll, M. Hoffman and M. Trodden about effective field theory, where the phantom model can be considered as a part or a range of a more fundamental theory \cite{Trodden} (see also \cite{Richarte}). Particularly, in these works it was shown that there exist a low-energy boundary such that the phantom field would be stable throughout the Universe's lifespan.

\par From the other hand, the analysis of the observation data obtained by different groups of researchers for the cosmological acceleraion and for the the \emph{barotrope coefficient} $w=p/\varepsilon$, related to it in the recent years, shows that, apparently, cosmology would be very hardly manageable without phantom fields. For instance, the SNIa data significantly favoring the ``phantom'' models and excludes the cosmological constant \cite{Tripathi}. The strict limitations can be obtained in combination with the observation data including the measurements of the Hubble parameter  $H(z)$ for various red shifts. When combining the standard rulers and standard clocks, the best conformity is being observed at $w_0 = -1.01 (+0.56 -0.31)$ \cite{Ma}. For the flat wCDM model the constant parameter of the dark energy's equation of state was measured $w = -1.013(+0.068 -0.073)$ \cite{Meyers}, see also \cite{Terlevich, Chavez}.

\par So, notwithstanding various difficulties, the cosmological models with the phantom scalar fields have the right to exist. From the theoretical perspective such models are very interesting and at the same time, insufficiently researched. In the more recent Authors' works the non-minimal theory of the scalar interaction based on the concept of the fundamental scalar charge has been being consequently developed both for classical and phantom scalar fields \cite{Ignat_12_1_Iz}, \cite{Ignat_12_2_Iz}, \cite{Ignat_12_3_Iz}. In particular, certain peculiarities of the phantom fields were revealed in these works, such as, for example, the peculiarities of the interparticle interaction. These researches were deepened for extension of the theory of the scalar fields including phantom ones to the sector of negative particle masses, degenerated Fermi-systems, conformally invariant interactions etc. \cite{Ignat_14_1_stfi,Ignat_Dima14_2_GC,Ignat_Agaf_Dima14_3_GC,Ignat_15_1_GC,Ignat_Agaf15_2_GC}. The mathematical models of the scalar fields that were constructed that way, were applied in the research of the cosmological evolution of the system of interacting particles and scalar fields of both classical and phantom types \cite{Ignat_Mih15_2_Iz}, \cite{Ignat_Agaf_Mih_Dima15_3_AST, Ignat_Agaf_16_3}. These researches revealed the unique properties of the cosmological evolution of plasma with interparticle phantom scalar interaction such as the existence of giant bursts of the cosmological acceleration, presence of plateau with constant acceleration and other anomalies due to which the behavior of cosmological models with the phantom scalar fields differ markedly from the models with the classical scalar field. In particular, the classification of behavioral types of cosmological models with interparticle phantom scalar field was performed and 4 essentially different models were selected in the works \cite{Ignat_Agaf_16_3} -- \cite{Ignat_Sasha_G&G}  In the same works it was also noted the possibility of Bose condensation of nonrelativistic scalar charged fermions at strong growth of  the scalar field's potential and consideration of this condensate in the capacity of the dark matter's component. It is important to note the circumstance that in the case of the phantom field with attraction in the course of cosmological evolution the values of the cosmological acceleration greater than 1 are achievable which corresponds to the  phantom state of matter as per generally accepted classification.
\par The cited above researches demonstrate the necessity to investigate more in details the classical and the phantom scalar fields with the self-action as  possible basis of the cosmological model of the early Universe. In the papers \cite{Ignat_16_5_Iz} -- \cite{Ignat_Agaf_2017_2_GC} it was carried out the preliminary qualitative analysis of the cosmological model based on the phantom scalar field with the self-action. In the given work we develop and elaborate the results of researches of the cosmological models based on the classical and the phantom scalar fields. As opposed to the works \cite{Ignat_14_1_stfi} -- \cite{Ignat_Agaf_Mih_Dima15_3_AST} we do not account the contribution of the ordinary matter i.e. we consider the free classical and phantom fields without a source.

\section{The Basic Relations of the Cosmological Model Based on the Single Scalar Field}
\subsection{The Field Equation}
The Lagrangian function of the scalar field $\Phi$ with the self-action has the following form:
\begin{equation}\label{Lagrange0}
L=\frac{1}{8\pi}\left(e_1 g^{ik}\Phi_{,i}\Phi_{,k}-2V(\Phi)\right),
\end{equation}
where it is
\begin{eqnarray}
\label{V(phi)}
V(\Phi)= -\frac{\alpha}{4}\left(\Phi^2-e_2\frac{m^2}{\alpha}\right)^2,
\end{eqnarray}
where $\alpha $ is the self-action constant, $m$ is the scalar field's mass of quanta; it is  $e_1=1$ for the classical scalar field, it is  $e_1=-1$ for the phantom scalar field. The potential energy $U$ is defined by the relation:
\begin{eqnarray}\label{potential}
4\pi U= V(\Phi)\equiv -\alpha\frac{\Phi^4}{4}+e_2 \frac{m^2\Phi^2}{2}-\frac{m^4}{4\alpha}\nonumber\\
\Rightarrow 4\pi U= -\alpha\frac{\Phi^4}{4}+e_2 \frac{m^2\Phi^2}{2}+\mathrm{Const},
 \end{eqnarray}
so that
\begin{eqnarray}\label{+-consts1}
U(-\alpha,-e_2,\pm\Phi)\equiv - U(\alpha,e_2,\pm\Phi).
\end{eqnarray}
The additive constant in the potential function can be dropped, therefore in the limit of the small constant of self-action we have the  ``massive term'' in the Lagrangian (\ref{Lagrange0})
\[U=e_2 \frac{m^2\Phi^2}{4\pi }, \quad \alpha\to 0.\]
Thus, in this sense, we can conditionally separate out the following cases;
\begin{enumerate}
\item $e_2=1$ is the scalar field with attraction;
\item $e_2=-1$ is the scalar field with repulsion.
\end{enumerate}

\begin{center}
\includegraphics[width=8cm]{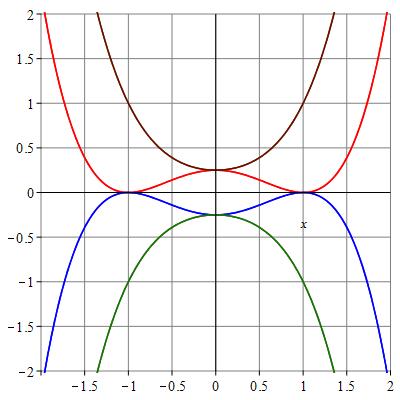}
\figcaption{\label{ris1}  The plots of the potential energy $4\pi U(\alpha,e_2,\Phi)$. The parabolas (dotted lines): the upper one  $4\pi U(-1,1,\Phi)$ and the lower one $4\pi U(1,-1,\Phi)$; and the parabolas with three extreme points (the solid lines): the upper $4\pi U(-1,-1,x)$ and the lower $4\pi U(1,1,x)$. The upper parabolas correspond to the attraction and the lower - to the repulsion.}
\end{center}

The tensor of the energy-momentum of the scalar field relative to the Lagrangian function (\ref{Lagrange0}) takes the standard form:
\begin{equation}\label{T_ik0}
T_{ik}=\frac{1}{8\pi}\bigl(2e_1\Phi_{,i}\Phi_{,k}-g_{ik}\Phi_{,j}\Phi^{,j}+2V(\Phi)g_{ik}\bigr).
\end{equation}
The equality to null of the covariant divergence of this tensor leads us to the equation of the free scalar field:
\begin{equation} \label{EqField0}
\square \Phi +V'(\Phi) =0.
\end{equation}
Since we can append an arbitrary constant to the Lagrange function \footnote{which leads to the renormalization of the cosmological constant}, we further omit the corresponding constant in the potential function where it leads to simplifications. Such a renormalization returns us back to the initial Lagrange function of the scalar field with the self-action which was used in the papers \cite{Ignat_16_5_Iz} -- \cite{Ignat_Agaf_2017_2_GC}, and which we will use further (see \cite{Ignat83_1}):
\begin{equation} \label{Lagrange}
L=\frac{1}{8\pi } \left(e_1g^{ik} \Phi _{,i} \Phi _{,k} - e_2 m^{2} \Phi ^{2} +\frac{\alpha }{2} \Phi ^{4} \right),
\end{equation}
The enery -- momentum reltive to the Lagrangian function (\ref{Lagrange}) is equal to
\begin{eqnarray}\label{T_ik}
T_{ik}=\frac{1}{8\pi}\bigl(2e_1\Phi_{,i}\Phi_{,k}-g_{ik}e_1\Phi_{,j}\Phi^{,j}+\nonumber\\
g_{ik}e_2 m^2\Phi^2-g_{ik}\frac{\alpha}{2}\Phi^4\bigr).
\end{eqnarray}
Using the Lagrangian function in the form (\ref{Lagrange}), let us obtain from~(\ref{EqField0}):
\begin{equation} \label{EqField1}
\Box\Phi+m^2_*\Phi=0,
\end{equation}
where $m_*$ is the effective mass of a scalar boson
\begin{equation}\label{m_*}
m^2_* \equiv e_1 (e_2 m^2-\alpha\Phi^2),
\end{equation}
which theoretically can be an imaginary value.

Let us write out the Einstein equation with the cosmological term\footnote{ We use the Planck system of units:
$G=c=\hbar =1$; the Ricci tensor is obtained by convolution of the first and the forth indices  $R_{ik}=R^j_{~ikj}$; the metric has the signature $(-1,-1,-1,+1)$.}
\begin{equation} \label{EqEinstein0}
R^{ik} -\frac{1}{2} Rg^{ik} =\lambda g^{ik} +8\pi T^{ik},
\end{equation}
where $\lambda\geq 0$ is the cosmological constant.
\subsection{The Equations of the Cosmological Model}
Let us write out the self-consistent system of equations of the cosmological model based on the free scalar field and space-flat Friedmann metric (\ref{EqField1}) -- (\ref{EqEinstein0}), supposing $\Phi=\Phi(t)$:
\begin{equation}\label{metric}
ds^{2} =dt^{2} -a^{2} (t)(dx^{2} +dy^{2} +dz^{2}).
\end{equation}
The mentioned system comprises of the single Einstein equation
\begin{equation}\label{EqEinstein}
3\frac{\dot{a}^{2}}{a^{2}} = e_1\dot{\Phi }^{2} +e_2 m^{2}\Phi^{2} -\frac{\alpha }{2} \Phi ^{4} +\lambda
\end{equation}
and the equation of the scalar field:
\begin{equation}\label{Eq_fild}
\ddot{\Phi }+3\frac{\dot{a}}{a} \dot{\Phi } +m_{*}^{2} \Phi =0,
\end{equation}
where $\dot(f)\equiv fd/dt$. Herewith the tensor of the energy-momentum (\ref{T_ik}) has a structure of the energy -- momentum tensor of the isotropic liquid with the following energy density and pressure:
\begin{eqnarray} \label{varepsilon}
\varepsilon =\frac{1}{8\pi } \left(e_1\dot{\Phi }^{2} +e_2 m^{2}\Phi^{2} -\frac{\alpha }{2} \Phi ^{4} \right);\\
\label{p}
p=\frac{1}{8\pi } \left(e_1\dot{\Phi }^{2} -e_2 m^{2}\Phi ^{2} +\frac{\alpha }{2} \Phi ^{4} \right),
\end{eqnarray}
so that:
\[\varepsilon +p=\frac{e_1}{4\pi }\dot{\Phi}^{2} .\]

\par Further we will need the values of 2 kinematic functions of the Friedmann Universe:
\begin{equation} \label{H(t)}
H(t)=\frac{\dot{a}}{a} \ge 0;{\rm \; \; }\Omega (t)=\frac{a\ddot{a}}{\dot{a}^{2} } \equiv 1+\frac{\dot{H}}{H^{2} }
\end{equation}
is the Hubble constant $H(t)$ and the invariant cosmological acceleration $\Omega(t)$, which is invariant and defined in the next way with a use of the \emph{barotrope coefficient} $\varkappa=p/\varepsilon$\footnote{In the commonly accepted notation $\varkappa=w$.}:
\begin{equation} \label{kappa}
\Omega= -\frac{1}{2}(1+3\varkappa).
\end{equation}
Differentiating (\ref{EqEinstein}) with an account of the definition (\ref{H(t)}) and the field equation (\ref{Eq_fild}) we find:
\begin{equation}\label{dotH}
\dot{H}=-e_1\dot{\Phi}^2.
\end{equation}
Thus, for the classical scalar fields it is $\dot{H}\leq0$, for the phantom scalar fields it is $\dot{H}\geq0$, therefore for the classical fields it is $\Omega\leq1$, and for the phantom it is $\Omega\geq1$.

\par Let us consider the possible particular cases. For the \emph{classical scalar field with attraction} ($e_1=e_2=1$) the system of equations (\ref{EqEinstein}) -- (\ref{Eq_fild}) takes the next form:
\begin{equation}\label{EqEinstein_clas}
3H^2 =\dot{\Phi }^{2} +m^{2}\Phi^{2} -\frac{\alpha }{2} \Phi ^{4} +\lambda,
\end{equation}
\begin{equation}\label{Eq_fild_clas}
\ddot{\Phi }+3H \dot{\Phi } +m^{2} \Phi -\alpha\Phi^3=0,
\end{equation}
where the energy density and pressure of the scalar field are equal to:
\begin{eqnarray} \label{varepsilon_clas}
\varepsilon =\frac{1}{8\pi}\left(\dot{\Phi }^{2} +m^{2}\Phi^{2} -\frac{\alpha}{2}\Phi ^{4}\right) ;\nonumber\\
\label{p_clas}
p=\frac{1}{8\pi}\left(\dot{\Phi }^{2} - m^{2}\Phi ^{2} +\frac{\alpha }{2} \Phi ^{4}\right).
\end{eqnarray}

\par For the  \emph{classical scalar field with repulsion} ($e_1=1; e_2=-1$; $m^2_*=-m^2-\alpha\Phi^2$) the system of equations (\ref{EqEinstein}) -- (\ref{Eq_fild}) takes the next form:
\begin{equation}\label{EqEinstein_clas1}
3H^2 =\dot{\Phi }^{2} -m^{2}\Phi^{2} -\frac{\alpha }{2} \Phi ^{4} +\lambda,
\end{equation}
\begin{equation}\label{Eq_fild_clas1}
\ddot{\Phi }+3H \dot{\Phi } -m^{2} \Phi -\alpha\Phi^3=0,
\end{equation}
where the energy density and pressure of the scalar field are equal to:
\begin{eqnarray} \label{varepsilon_clas1}
\varepsilon =\frac{1}{8\pi}\left(\dot{\Phi }^{2} -m^{2}\Phi^{2} -\frac{\alpha}{2}\Phi ^{4}\right) ;\nonumber\\
\label{p_clas1}
p=\frac{1}{8\pi}\left(\dot{\Phi }^{2} + m^{2}\Phi ^{2} +\frac{\alpha }{2} \Phi ^{4}\right).
\end{eqnarray}
%

\par For the \emph{phantom scalar field with attraction} ($e_1=-1; e_2=1$, $m^2_*=-m^2-\alpha\Phi^2$) it is:
\begin{equation}\label{EqEinstein_fan1}
3H^2 =-\dot{\Phi }^{2} +m^{2}\Phi^{2} -\frac{\alpha }{2} \Phi ^{4} +\lambda,
\end{equation}
\begin{equation}\label{Eq_fild_fan1}
\ddot{\Phi }+3H\dot{\Phi }-m^{2} \Phi +\alpha\Phi^3=0,
\end{equation}
\begin{eqnarray} \label{varepsilon_clas}
\varepsilon =\frac{1}{8\pi}\left(-\dot{\Phi }^{2} +m^{2}\Phi^{2} -\frac{\alpha}{2}\Phi ^{4}\right) ;\nonumber\\
\label{p_clas}
p=\frac{1}{8\pi}\left(-\dot{\Phi }^{2} -m^{2}\Phi ^{2} +\frac{\alpha }{2} \Phi ^{4}\right).
\end{eqnarray}

\par For the \emph{phantom scalar field with repulsion} ($e_1=e_2=-1$; $m^2_*=m^2-\alpha\Phi^2$) it is:
\begin{equation}\label{EqEinstein_fan}
3H^2 =-\dot{\Phi }^{2} -m^{2}\Phi^{2}-\frac{\alpha }{2} \Phi ^{4} +\lambda,
\end{equation}
\begin{equation}\label{Eq_fild_fan}
\ddot{\Phi }+3H \dot{\Phi } +m^{2} \Phi +\alpha\Phi^3=0,
\end{equation}
\begin{eqnarray} \label{varepsilon_clas}
\varepsilon =\frac{1}{8\pi}\left(-\dot{\Phi }^{2} -m^{2}\Phi^{2} -\frac{\alpha}{2}\Phi ^{4}\right) ;\nonumber\\
\label{p_clas}
p=\frac{1}{8\pi}\left(-\dot{\Phi }^{2} +m^{2}\Phi ^{2} +\frac{\alpha }{2} \Phi ^{4}\right).
\end{eqnarray}

\section{The Qualitative Analysis}
\subsection{The Reduction of the System of Equations to the Normal Form}
Using the fact that the Hubble constant can be expressed from the Einstein equation (\ref{EqEinstein}) through the functions $\Phi,\dot{\Phi}$, proceeding to the dimensionless \emph{Compton time}:
\begin{equation}\label{tau}
mt=\tau;\quad (m\not\equiv0)
\end{equation}
and carrying out a standard substitution of variables $\Phi'=Z(\tau)$ ($f'\equiv df/d\tau$), let us reduce the Einstein equation (\ref{EqEinstein}) to the dimensionless form:
\begin{equation}\label{Hm}
H'\ \!\!^2_m=\frac{1}{3}\left[e_1 Z^{2} +e_2\Phi^{2} -\frac{\alpha_m}{2}\Phi ^{4} +\lambda_m\right],
\end{equation}
and the field equation (\ref{Eq_fild}) - to the form of normal autonomous system of ordinary differential equations on the plane~$\{\Phi,Z\}$:
\begin{eqnarray} \label{DynSys}
\Phi ' &=& Z;\nonumber\\
Z' &=& \displaystyle-\sqrt{3}Z \sqrt{e_1 Z^{2} + e_2 \Phi ^{2} -\frac{\alpha _{m} }{2} \Phi ^{4} + \lambda _{m}} \nonumber\\ & & \displaystyle - e_1e_2 \Phi +e_1\alpha _{m} \Phi ^{3} ,
\end{eqnarray}
where the following denotations are introduced:
\[\lambda _{m} \equiv \frac{\lambda }{m^{2}};\; \alpha _{m} \equiv \frac{\alpha }{m^{2} } . \]
Here it is:
\begin{equation} \label{GrindEQ__12_}
\frac{a'}{a}\equiv \Lambda'=H_m\equiv \frac{H}{m};\quad \Omega =\frac{aa''}{a'^{2} } \equiv 1+\frac{h'}{h^{2} } ,
\end{equation}
where
\begin{equation}\label{Lambda}
\Lambda=\ln a(\tau).
\end{equation}
Let us note that all the values $\Phi, Z, H_m, \alpha_m,\Omega, \tau$ in this denotation are dimensionless; the time $\tau$ is changed at that in the Compton scale.

Thus, we have the autonomous 2-dimensional dynamic system on the phase plane $\{ \Phi ,Z\} $. To reduce it to the standard notation of the qualitative theory of differential equations (see e.g.,  [15]) let us assume:
\begin{eqnarray}\label{GrindEQ__13_}
\Phi =x;\;Z=y;&\nonumber\\
P(x,y)=y;&\nonumber\\
Q(x,y)=& \displaystyle-\sqrt{3} y\sqrt{ e_1 y^{2} + e_2 x^{2} -\frac{\alpha _{m} }{2} x^{4} + \lambda _{m}}\nonumber\\
&\displaystyle -e_1e_2 x +e_1\alpha _{m} x^{3}.
\end{eqnarray}
The corresponding normal system of equations in the standard notation has the following form:
\begin{equation} \label{DynSysPQ}
x'=P(x,y);{\rm \; \; \; }y'=Q(x,y) .
\end{equation}

\subsection{The Accessible Regions of Motion on the Phase Plane}\label{Top_1}
For the system of differential equations \eqref{DynSys} (or (\ref{DynSysPQ})) to have real solution, it is necessary that the following inequality is fulfilled (see Fig. \ref{ris2} -- \ref{ris4}):
\begin{equation} \label{GrindEQ__15_}
e_1 y^{2} + e_2 x^{2} -\frac{\alpha _{m} }{2} x^{4} + \lambda _{m} \ge 0.
\end{equation}
The boundary of the range (\ref{GrindEQ__15_}) is defined by the algebraic equation of the 4th order:
\begin{equation}\label{Eq_xy4}
e_1 y^{2} + e_2 x^{2} -\frac{\alpha _{m} }{2} x^{4} + \lambda _{m}=0.
\end{equation}
Introducing the variables,
\begin{equation}\label{eta,xi}
u=x^2-\frac{e_2}{\alpha_m};\; v=y^2+e_1\left(\frac{1}{2\alpha_m}+\lambda\right),
\end{equation}
the inequality (\ref{GrindEQ__15_}) can be transformed to the next inequalities:
\begin{equation}\label{noneq_par}
\begin{array}{ll}
\displaystyle u^2\leq \frac{2e_1}{\alpha_m}v, & \alpha>0;\\[12pt]
\displaystyle u^2\geq  \frac{2e_1}{\alpha_m}v, & \alpha<0.
\end{array}
\end{equation}
and the equation (\ref{Eq_xy4}) -- to the canonical equation of the parabola with an axis $v$:
\begin{equation}\label{par_xi,eta}
u^2=\frac{2e_1}{\alpha_m}v.
\end{equation}
Let us notice that for the classical field ($e_1=1$) in the case when $\alpha>0$ it is always $v>0$, therefore the variable $u$ can take the values only in the bounded interval.

Let us also notice that the curve of the 4th order (\ref{Eq_xy4}) corresponds to the null value of the Hubble constant i.e null expansion velocity.
From the other hand, if we consider the sum of the energy of the scalar field and the cosmological term's contribution as the effective energy of the matter
\begin{eqnarray}\label{E_ef}
\varepsilon_{eff}=\varepsilon+\frac{1}{8\pi}\lambda\equiv=m^2\varepsilon_{m}\\
=\frac{m^2}{8\pi}\left(e_1 y^{2} + e_2 x^{2} -\frac{\alpha _{m} }{2} x^{4} + \lambda _{m}\right) ,\nonumber
\end{eqnarray}
the curve (\ref{Eq_xy4}) will be corresponded by the null effective energy $\varepsilon_{eff}=0$.

\end{multicols}
\ruleup
\begin{center}
\includegraphics[width=16cm]{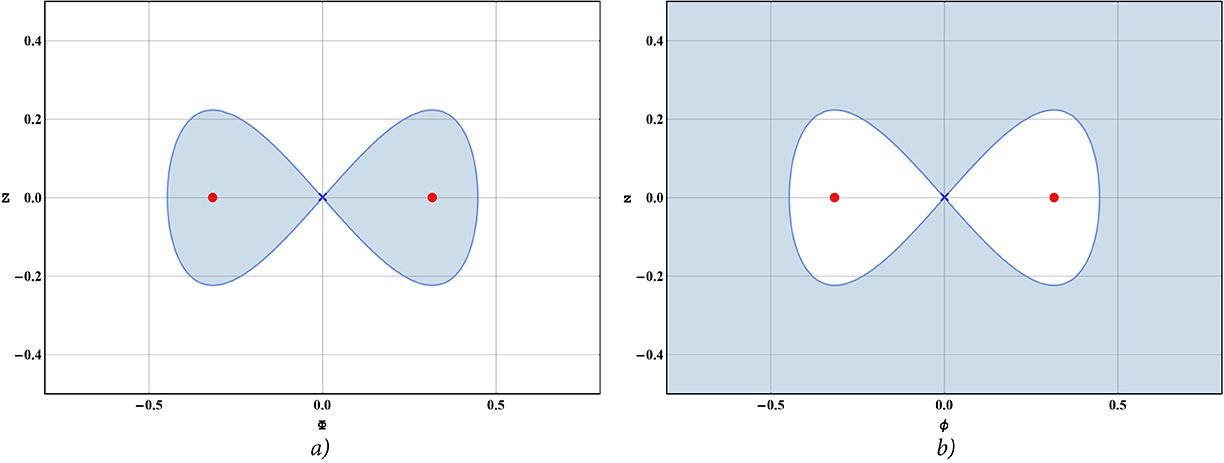}
\figcaption{\label{ris2} The prohibited ranges of motion of the dynamic system. \textbf{a)} The classical scalar field ($e_1=1,\ e_2=-1,\ \alpha=-10,\lambda =0$); \textbf{ b)} The phantom scalar field ($e_1=-1,\ e_2=1,\ \alpha=10,\lambda =0$). The motion is only possible in the unpainted ranges. Here and further the attractive centers are marked with red dots and the saddle points are marked with blue crosses.}
\end{center}

\begin{center}
\includegraphics[width=16cm]{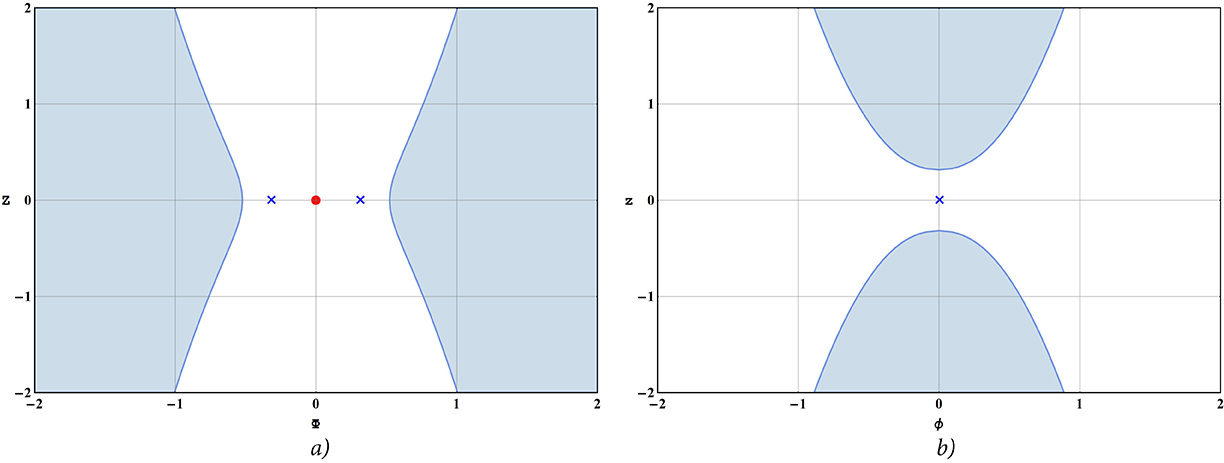}
\figcaption{\label{ris3} The prohibited ranges of motion of the dynamic system. \textbf{a)} The classical scalar field ($e_1=1,\ e_2=1,\ \alpha=10,\lambda =0.1$); \textbf{ b)} The phantom scalar field ($e_1=-1,\ e_2=1,\ \alpha=-10,\lambda =0$). The motion is  possible only in the unpainted ranges.}
\end{center}

\begin{center}
\includegraphics[width=16cm]{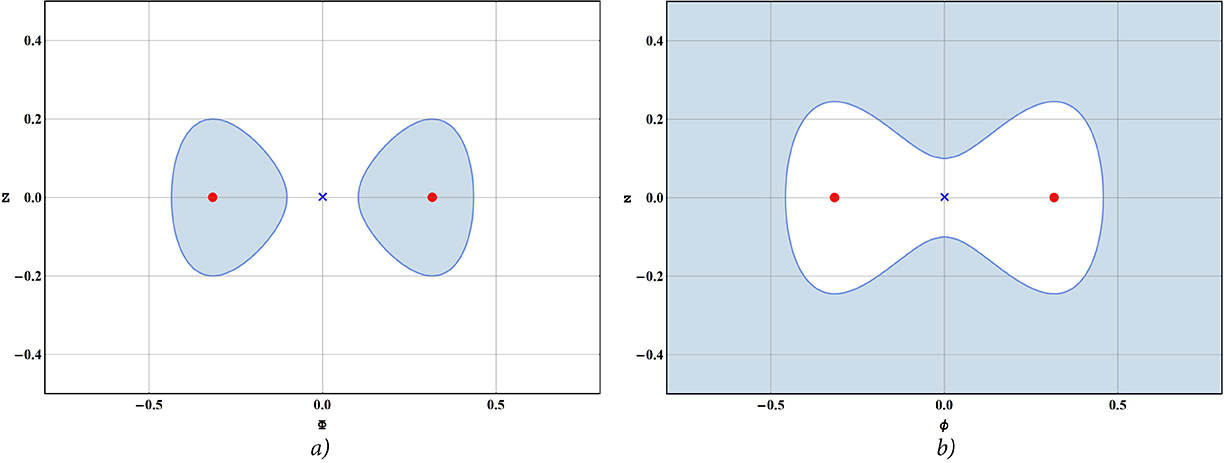}
\figcaption{\label{ris4} The prohibited ranges of motion of the dynamic system. \textbf{a)} The classical scalar field ($e_1=1,\ e_2=-1,\ \alpha=-10,\lambda =0.01$);  \textbf{b)} The phantom scalar field ($e_1=-1,\ e_2=1,\ \alpha=10,\lambda =0.01$). The motion is  possible only in the unpainted ranges.}
\end{center}

\ruledown

\begin{multicols}{2}

At $\lambda=0$ the external range, which is limited by this curve, is an area of allowable values of the dynamic values for the classical field (Fig. \ref{ris2} a); the admitted values of the dynamic variables for the phantom field, on the contrary, lie in the internal area (Fig. \ref{ris2} b). The invertibility property disappears at $\lambda\not=0$ and the split of the area of allowable values of the dynamic variables into two symmetrical ranges become possible (Fig. \ref{ris3}, \ref{ris4}). Thus, the nonlinear scalar fields, both classical nd phantom ones, are characterized by the violation of the simple connectedness of the phase space and appearance of the white areas, inaccessible for the dynamic system. This fact can lead to the fundamental differences of the asymptotic behavior of the cosmological models based on nonlinear scalar fields.

\subsection{The Singular Points of the Dynamic System}
The singular points of the dynamic system are defined by means of the equations (see, e.g.  \cite{Bogoyav}, \cite{Bautin}):
\begin{equation} \label{GrindEQ__16_}
M:{\rm \; \; }P(x,y)=0;{\rm \; }Q(x,y)=0.
\end{equation}
Since according to (\ref{GrindEQ__13_}) and (\ref{DynSysPQ}) in any of the singular points of the dynamic system it is always $Z\equiv y=0$, let us obtain the following equation in order to find the solutions:
\begin{equation}\label{Eq_Sing}
x(e_2-\alpha_m x^2)=0.
\end{equation}
Let us note that at the singular points of the dynamic system, according to (\ref{dotH}) it is:
\begin{equation}\label{dotH=0}
\dot{H}=0,
\end{equation}
therefore all the singular points of the dynamic system lie on the inflationary trajectory $\Omega=1$.

Then, at any $\alpha _{m} $ and $\lambda _{m} \ge 0$ the system of the algebraic equations \eqref{GrindEQ__16_} always has the trivial solution
\begin{equation} \label{Phi=0Z=0}
x=0;y=0{\rm \; }\Rightarrow M_{0} (0,0),
\end{equation}
and at $e_2\alpha>0$ -- two more symmetrical solutions:
\begin{equation} \label{Phi_pmZ0}
x=x_\pm=\pm \frac{1}{\sqrt{e_2\alpha_{m}}};\ y=0\quad \Rightarrow M_\pm(x_\pm,0).
\end{equation}
Substituting the solutions \eqref{Phi=0Z=0} and \eqref{Phi_pmZ0} in the condition (\ref{GrindEQ__15_}), we find the necessary condition of the solutions' reality in the singular points \eqref{Phi=0Z=0} and \eqref{Phi_pmZ0}:
\begin{equation} \label{ReM}
(\ref{Phi=0Z=0})\rightarrow \lambda_m\ge 0;\;(\ref{Phi_pmZ0})\rightarrow\lambda_m +\frac{1}{2\alpha_m}\ge 0.
\end{equation}

\subsection{The Charachteristic Equation and the Qualitative Analysis in the Neighborhood of the Null Singular Point}
Let us calculate the derivatives of the functions \eqref{GrindEQ__13_} at null singular point $M_0$ \eqref{GrindEQ__16_} at $\lambda _{m} \ge 0$ :
\[
\begin{array}{ll}
\displaystyle\left.\frac{\partial P}{\partial x} \right|_{M_0} =0;&\hspace{1cm}\displaystyle \left.\frac{\partial P}{\partial y}\right|_{M_0}=1;\\
\displaystyle\left.\frac{\partial Q}{\partial x} \right|_{M_0 }=-e_1e_2;&\hspace{1cm} \left. \displaystyle\frac{\partial Q}{\partial y} \right|_{M_0} =\displaystyle -\sqrt{3\lambda_m}.
\end{array}
\]
The matrix of the dynamic system at null point is:
\[A_{M_0}=\left(
\begin{array}{cc}
0 & {1} \\[12pt]
{-e_1e_2} & 0 -\sqrt{3\lambda_{m} }
\end{array}
\right).\]
Its determinant is equal to:
\begin{equation}\label{Det0}
\Delta_0=\mathrm{det}(A)_{M0}=e_1e_2.
\end{equation}
Thus, we get the characteristic equation and its roots $k_\pm$:
\begin{eqnarray} \label{x0kpm}
\left|
\begin{array}{cc}
-k & {1} \\[12pt]
{-e_1e_2} & {-k -\sqrt{3\lambda_{m} } }
\end{array}
\right|=0\Rightarrow\nonumber\\
k_{\pm} =-\frac{\sqrt{3\lambda _{m} } }{2} \pm \frac{\sqrt{3\lambda_{m} - 4e_1e_2} }{2}.
\end{eqnarray}
As is well known, the product of the eigen-values of the matrix is equal to its determinant:
\begin{equation}\label{k1k2}
k_1 k_2=\mathrm{det}(A).
\end{equation}
Therefore the signs of the eigenvalues at null as a consequence of (\ref{Det0}) is defined by the index $e_2$.
It is also easy to see that the eigenvalues at the null singular point do not at all depend neither on the constant of the self-action nor on the character of the field
(classical/phantom), i.e. do not depend on the value of the parameter $e_1$. Let us consider in details all the possible cases.

\subsubsection{The Classical Filed with the Attraction: $e_1=e_2=1$ and the Phantom Field with the Repulsion: $e_1=e_2=-1$}
These two cases are identical at the null singular point $M_0$. Here the 4 different situations are possible:\\
\noindent 1a.~ \emph{The Case of the Null Value of the Cosmological Constant}
\begin{equation}\label{l=0}
\lambda=0 \rightarrow k=\pm i.
\end{equation}
Since the eigenvalues turned to be purely imaginary ones, then \emph{the single singular point} (\ref{Phi=0Z=0}) of the dynamic system (\ref{DynSys}) is its center (see \cite{Bogoyav}). In this case the phase trajectory of the dynamic system is winding onto this center making an infinite number of turns.\\
\noindent 1b.~ \emph{The Case of the Small Value of the Cosmological Term}:
\begin{equation}\label{Lambda<4/3}
0<\lambda_m<\frac{4}{3}
\end{equation}
-- we have two complex conjugate eigenvalues and it is
\begin{equation}\label{Re<0}
\Re(k)=-\frac{\sqrt{3\lambda_m}}{2}<0.
\end{equation}
In this case in accordance with the qualitative theory of the differential equations the point $M_0$ (\ref{Phi=0Z=0}) is \emph{the attractive focus}, --
all the phase trajectories of the dynamic system at $\tau\to+\infty$ are the twisting spirals winding onto the singular point, making an infinite number of turns at that. This case practically coincides with the previous one (Fig. \ref{ris5}).\\

\begin{center}
\includegraphics[width=8cm]{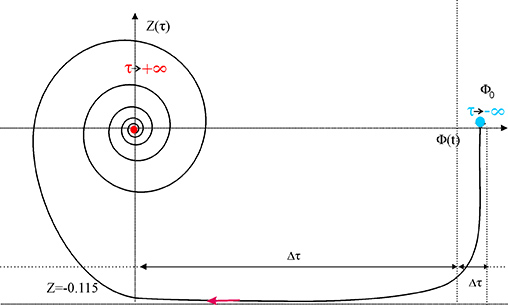}
\figcaption{\label{ris5}  The qualitative view of the phase trajectory of the dynamic system (\ref{DynSys}) for the classical scalar field at $\lambda_m<4/3$. On this figure $\Delta\tau$ is the characteristic fall time of the potential's speed change till the bottom of the plot, $Z_0\approx-0.115$, $\Delta t$ -- the characteristic fall time of the potential's value with the constant speed $\Phi'\approx Z_0$. After this time instant the phase trajectory starts to wind onto the null attractive focus/center. The number of turns of the spiral at that is infinite.}
\end{center}

\begin{center}
\includegraphics[width=8cm]{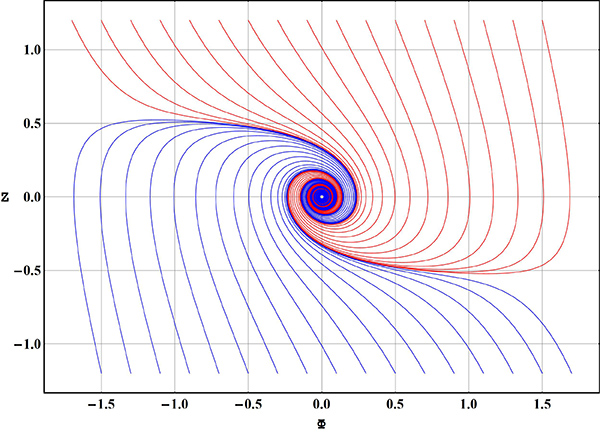}
\figcaption{\label{ris6} The phase portrait of the system (\ref{DynSys}) for the classical scalar field without the self-action in the case $e_1=1; e_2=1; \alpha_m=0$; $\lambda_m=0$}
\end{center}

\noindent 1c.~ \emph{The Case of the Large Value of the Cosmological Term}:
\begin{equation}\label{Lambda>4/3}
\lambda_m>\frac{4}{3}
\end{equation}
-- then we have two different real eigenvalues  $k_1\not=\lambda_2$, $k_1<0,k_2<0$, which are negative according to (\ref{x0kpm}). In this case the singular point is \emph{a stable attractive node}. At $\tau\to+\infty$ all the phase trajectories of the dynamic system come to the singular point and all the trajectories apart from two exceptional ones, when coming to the singular point, are tangent to the eigenvector $\mathbf{u}_1$, which corresponds to the minimal in magnitude eigenvalue, i.e., $k_1$. Two exceptional trajectories are tangent to the second eigenvector $\mathbf{u}_2$. The mentioned eigenvectors are equal to:
\begin{equation}\label{u1u2}
\mathbf{u}_1=(1,k_1);\quad \mathbf{u}_1=(1,k_2).
\end{equation}
The angle $\phi$ between the eigenvectors is defined by the relation:
\begin{equation}\label{alpha}
\cos\phi\equiv \frac{\mathbf{u}_1\mathbf{u}_2}{\sqrt{\mathbf{u}^2_1\mathbf{u}^2_2}}=\sqrt{\frac{4}{3\lambda_m}}<1\Rightarrow \phi>0.
\end{equation}
At very big values of $\lambda_m$ the angle between the vectors tends to $\pi/2$, at $\lambda_m\to4/3$ the angle tends to null.\\
\noindent 1d. \emph{The degenerated case}:
\begin{equation}\label{Lambda=4/3}
\lambda_m=\frac{4}{3}
\end{equation}
-- This case piratically coincides with the previous one with the only difference that it takes into account the fact that all the trajectories come to the singular point tangent to the single eigenvector -- this corresponds to the mentioned above boundary case $\phi\to0$.
\subsubsection{The Classical Field with the Repulsion: $e_1=1; e_2=-1$ or the Phantom Field with the Attraction: $e_1=-1; e_2=1$}
In this case independently of the field's character (classical/phantom) according to (\ref{x0kpm}) we have 2 real eigenvalues with opposite signs. Thus, the null singular point (\ref{Phi=0Z=0}) in this case is a {\it saddle}.
\subsection{The Characteristic Equation and the Qualitative Analysis in the Neighborhood of the Non-Zero Singular Point}
Let us calculate the derivatives of the fucntions \eqref{GrindEQ__13_} at non-zero singular points $M_\pm(x_\pm,0)$ \eqref{Phi_pmZ0} at $\lambda _{m} \ge 0$:
\[
\begin{array}{ll}
\displaystyle\left.\frac{\partial P}{\partial x} \right|_{M_\pm} =0;&\hspace{1cm}\displaystyle \left.\frac{\partial P}{\partial y}\right|_{M_\pm}=1;\\
\displaystyle\left.\frac{\partial Q}{\partial x} \right|_{M_\pm }=2e_1e_2;&\hspace{1cm} \left. \displaystyle\frac{\partial Q}{\partial y} \right|_{M_\pm} =\displaystyle -\sqrt{3}\sqrt{\lambda_m+\frac{1}{2\alpha_m}}.
\end{array}
\]
Let us notice that as a result of the condition (\ref{ReM}), the values of the derivatives are real. The determinant of the dynamic system's matrix is equal to:
\begin{equation}\label{Detpm}
\Delta_\pm=\mathrm{det}(A)_{M_\pm}=-2e_1e_2.
\end{equation}
This, we obtain the roots of the characteristic equation $k_\pm$, which coincide for the symmetrical points $M_\pm$:
\begin{eqnarray} \label{kpm}
k_{\pm } =\frac{\sqrt{3}}{2}\left[-\sqrt{\lambda_m+\frac{1}{2\alpha_m}}\pm
\sqrt{\lambda_m+\frac{1}{2\alpha_m}+\frac{8e_1e_2}{3}}
\right],
\end{eqnarray}
so that:
\begin{equation}\label{k1k2pm}
k_1 k_2=-2e_1e_2.
\end{equation}
\subsubsection{The Classical Field with Attraction: $e_1=e_2=1$ or the Phantom Fields with Repulsion: $e_1=e_2=-1$}
In consequence of \eqref{ReM} the radicand in the first term \eqref{kpm} is strictly greater then null. Moreover, as a result of $e_1e_2=+1$ the radicand in the second term \eqref{kpm} is greater than the radicand in the first term \eqref{kpm}, therefore both eigenvalues are real and opposite by sign. Thus, the points $M_\pm$ in this case are \emph{instable saddle points} at any direction of the time.

\subsubsection{The Classical Field with  Repulsion: $e_1=1; e_2=-1$ or the Phantom Fields with  Attraction: $e_1=-1; e_2=1$}

In consequence of \eqref{ReM} the radicand in the first term \eqref{kpm} is strictly greater than null. Moreover, as a result of $e_1e_2=-1$ the radicand in the second term \eqref{kpm} is less than the radicand in the firs term \eqref{kpm} and, in general, can be negative.
Thus, there are three possible cases:\\
1) $\lambda_m+1/2\alpha_m-8/3>0$ -- both eiegenvalues are real and negative. In this case the solution contains \textit{two symmetrical attractive (stable) nondegenerated nodes}. All phase trajectories in the neighborhood of such singular points at $t\to\infty$ come into these points and all of them, apart from two exceptional ones, are tangent to the eigenvector of the minimal length.
\noindent
2) $\lambda_m+1/2\alpha_m-8/3=0$ -- both eigenvalues are negative and equal to each other. In this case the solution contains \textit{two symmetrical nondegenerated nodes}, which are \emph{the bifuraction points} of the dynamic system.\\
3) $\lambda_m+1/2\alpha_m-8/3<0$ -- both eigenvalues are complex conjugate and their real parts are negative. In this case the solution contains\textit{two symmetrical attractive focuses}.

In the case of two symmetrical focuses it is easy to find the limiting value of $H_m(\infty)$, to which the Hubble constant tends to at $t \to \infty$. Substituting the coordinates of the focuses $M_{\pm}(\pm1/\sqrt{e_2\alpha}, 0)$ into the equation (\ref{Hm}), we obtain:
\begin{equation}
H_m(\infty) = \sqrt{\frac{1}{3}\left( \lambda_m - \frac{1}{2\alpha_m} \right)}.
\end{equation}

\section{The Numerical Simulation of the Cosmological Evolution in the Case of the Model with the Single Field for Typical Cases}
The results of the qualitative researches were confirmed and the fact of the asymptotic behavior of the trajectories at infinity was established as a consequence of the numerical simulation. Let us consider some examples.

\begin{center}
\includegraphics[width=8cm]{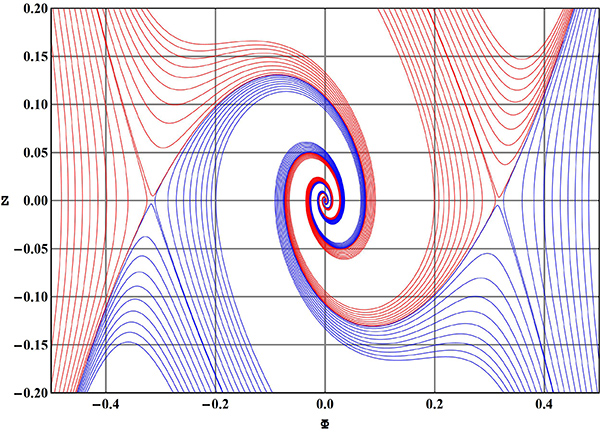}
\figcaption{\label{ris7}  The phase trajectories of the system (\ref{GrindEQ__13_}) for the classical field are: $e_1=1$, $e_2=1$, $\alpha_m=10$, $\lambda_m=0.1$.
The permitted area is shown on Fig. \ref{ris3}a.}
\end{center}

\begin{center}
\includegraphics[width=8cm]{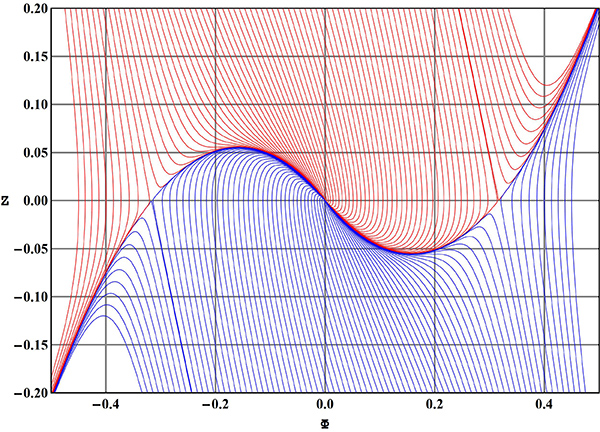}
\figcaption{\label{ris8}  The phase trajectories of the system (\ref{GrindEQ__13_}) for the classical field: $e_1=1$, $e_2=1$, $\alpha_m=10$, $\lambda_m=1.5$.}
\end{center}

\subsection{The Case of the Pair of the Saddle Singular Points ($e_1=e_2=1$; $e_1=e_2=-1$)}
In the real part of the solution (\ref{ReM}) in the case of the similar values of the parameters $e_1$ and $e_2$ the system has three singular points: a pair of the symmetrical saddle points with the coordinates $M_\pm(\displaystyle\pm \frac{1}{\sqrt{e_2\alpha_{m}}},0)$ and a null singular point $M_0(0,0)$. The figure \ref{ris9} illustrates the topological structure of the phase trajectories with the attractive focus in the origin of coordinates, the figure  \ref{ris10} illustrates the phase trajectories with the attractive center in the origin of coordinates.

\begin{center}
\includegraphics[width=8cm]{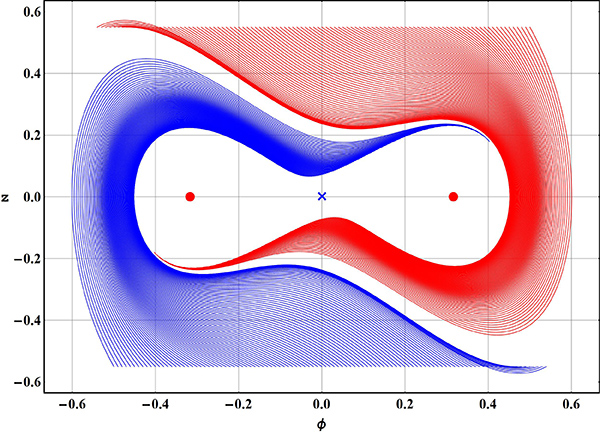}
\figcaption{\label{ris9}  The phase trajectories of the system (\ref{GrindEQ__13_}) for the classical field: $e_1=1$, $e_2=-1$, $\alpha_m=-10$, $\lambda_m=0$. The permitted area is shown on the Fig. \ref{ris2}a.}
\end{center}

\subsection{The Case of the Null Saddle Singular Point ($e_1=1$, $e_2=-1$; $e_1=-1$, $e_2=1$)}
If the conditions of the reality of the solution for the classical and phantom scalar fields in the case of different signs of the parameters $e_1$ and $e_2$ the system (\ref{GrindEQ__13_}) has the next singular points: a null saddle singular point and a pair of symmterical points, which character is defined by the parameters  $e_1$, $\alpha_m$, $\lambda_m$. For the numerical integration of the dynamic system (\ref{GrindEQ__13_}) it is required to select the initial values of the function $\Phi$, $Z$, satisfying the condition (\ref{GrindEQ__15_}). The figure \ref{ris9} illustrates the phase trajectories for the classical scalar field. It can be seen, that the ranges in the neighborhood of the symmetrical attractive focuses do not meet the condition (\ref{GrindEQ__15_}), therefore the trajectories having the origin in the $\tau=-\infty$ (the pole), finish on the boundary of this range\footnote{see below}.
The figure \ref{ris10} shows the phase portrait of the system for the phantom scalar field with the same parameter $\lambda_m$. It can be seen that the ranges far off the symmetrical attractive focuses are now unaccessible. In the given example $\lambda_m=0$ and the permitted area are now split by the directrices of the saddle point into two ranges each of which contains a family of the phase trajectories.

For the nonzero values of the parameter $\lambda_m$ the permitted area contains phase trajectories, part of which transitions from the neighborhood of one of the focuses to the neighborhood of the another one (Fig.~\ref{ris11}).
\begin{center}
\includegraphics[width=8cm]{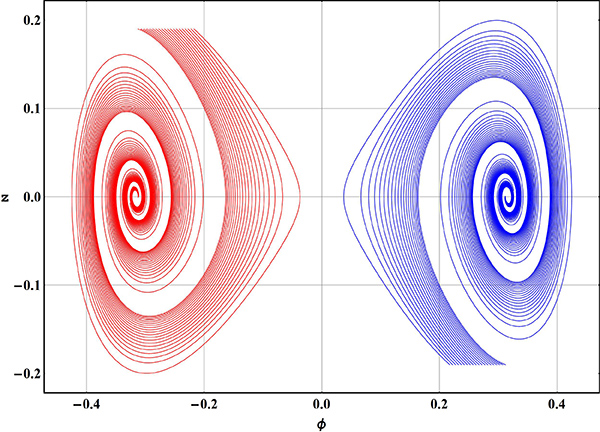}
\figcaption{\label{ris10}  The phase trajectories of the system (\ref{GrindEQ__13_}) for the phantom field: $e_1=-1$, $e_2=1$, $\alpha_m=10$, $\lambda_m=0$.}
\end{center}
\begin{center}
\includegraphics[width=8cm]{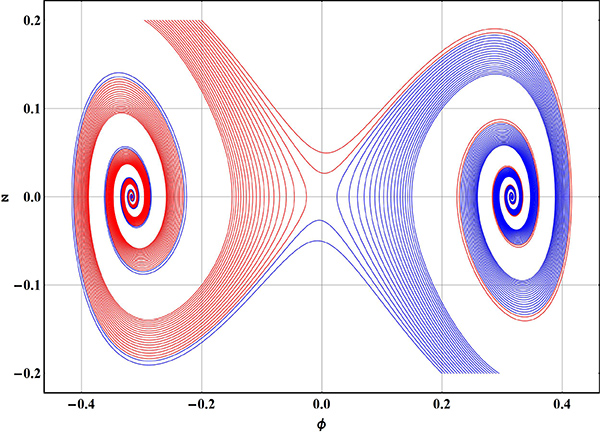}
\figcaption{\label{ris11}  The phase trajectories of the system (\ref{GrindEQ__13_}) for the phantom field: $e_1=-1$, $e_2=1$, $\alpha_m=10$, $\lambda_m=0.01$.}
\end{center}
At further increase of the parameter $\lambda_m$, the trajectories in the neighborhood of the symmetrical focuses are starting to bend (Fig. \ref{ris13}). The figure \ref{ris14} shows the case of the bifurcation of the dynamic system on change of the character of the symmetrical singular points from the attractive focus to the attractive centers, corresponding to the value $\lambda_m=2.7167$.
\begin{center}
\includegraphics[width=8cm]{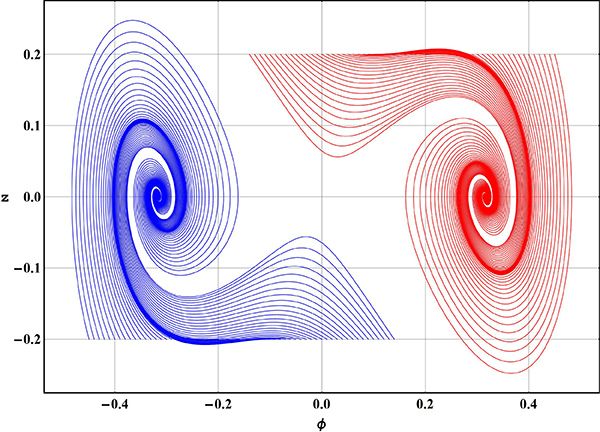}
\figcaption{\label{ris12}  The phase trajectories of the system (\ref{GrindEQ__13_}) for the phantom field: $e_1=-1$, $e_2=1$, $\alpha_m=10$, $\lambda_m=0.01$.}
\end{center}
\begin{center}
\includegraphics[width=8cm]{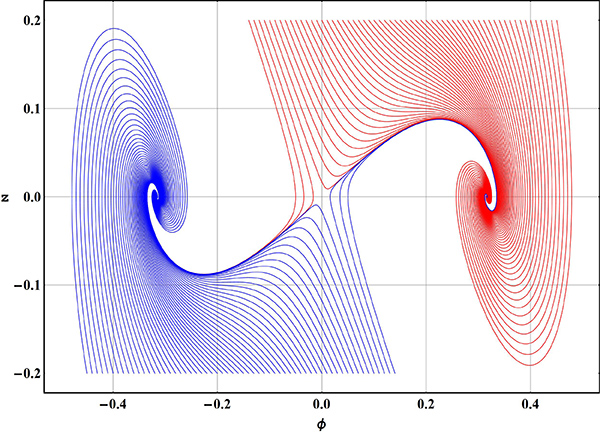}
\figcaption{\label{ris13}  The phase trajectories of the system (\ref{GrindEQ__13_}) for the phantom field: $e_1=-1$, $e_2=1$, $\alpha_m=10$, $\lambda_m=0.01$.}
\end{center}
\begin{center}
\includegraphics[width=8cm]{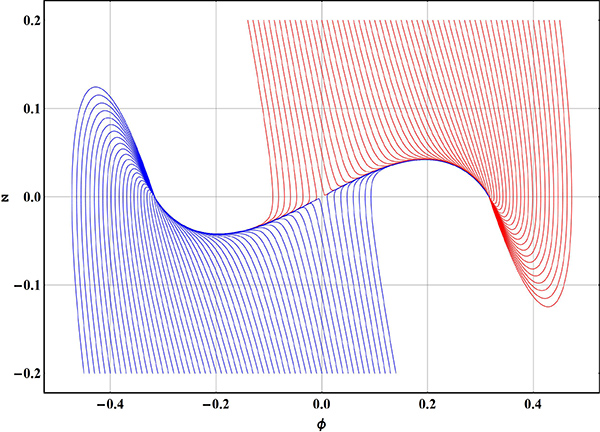}
\figcaption{\label{ris14}  The phase trajectories of the system (\ref{GrindEQ__13_}) for the phantom field: $e_1=-1$, $e_2=1$, $\alpha_m=10$, $\lambda_m=2.7167$. The bifurcation.}
\end{center}

\section{The Behavior of the Trajectories at Infinity: the Projection of the Phase Trajectories on the Poincare Sphere}

For the more deep research of the behavior of the dynamic system's trajectories at infinity let us construct the projections of the phase trajectories on the Poincare sphere. Applying the Poincare transformations \cite{Bogoyav,Bautin}
\begin{equation}\label{TransPouncare}
\xi = \frac{1}{\Phi},\;  \eta = \frac{Z}{\Phi}
\end{equation}
to the dynamic system (\ref{GrindEQ__13_}), we find the following relations in the new variables $u, v$:
\begin{eqnarray} \label{system_uv}
P^*(\xi,\eta) &=& -\eta\xi^3;\nonumber\\
Q^*(\xi,\eta) &=& \displaystyle-\sqrt{3} \;\eta\;\sqrt{ e_1 \eta^2 \xi^2 + e_2 \xi^2 -\frac{\alpha _{m} }{2} + \lambda _{m} \xi^4 }\nonumber\\
&&\displaystyle - e_1e_2 \xi^2 +e_1\alpha _{m} - \eta^2 \xi^2 .
\end{eqnarray}
The singular points at the equator of the Poincare sphere are defined from the conditions: $P^*(0,\eta) = 0$, $Q^*(0,\eta) = 0$. Thus, we find
\begin{equation}
\xi = 0;\; \eta_\infty = {\rm sgn}(\alpha_m)\sqrt{\frac{2}{3}}\sqrt{-e_1\alpha_m} \quad\Rightarrow\quad M_\infty(0,\eta_\infty).
\end{equation}
\begin{center}
\includegraphics[width=8cm]{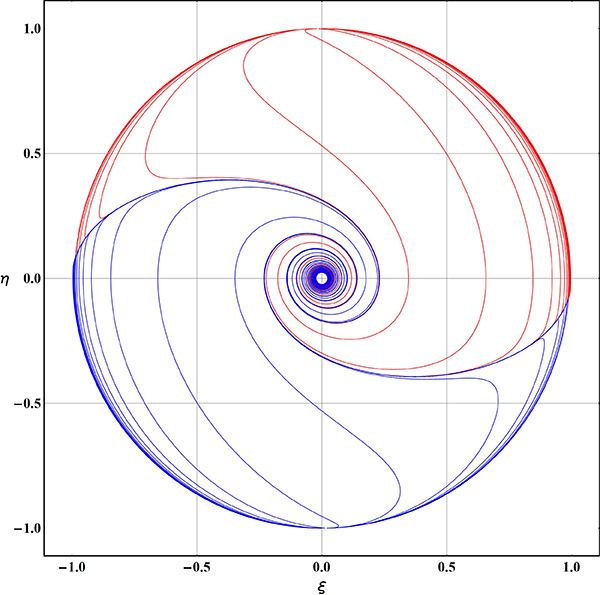}
\figcaption{\label{ris15} The phase trajectories of the system (\ref{GrindEQ__13_}) on the Poincare sphere for the classical field without a self-action at $e_1=1$, $e_2=1$, $\alpha_m=0$, $\lambda_m=0$.}
\end{center}
\begin{center}
\includegraphics[width=8cm]{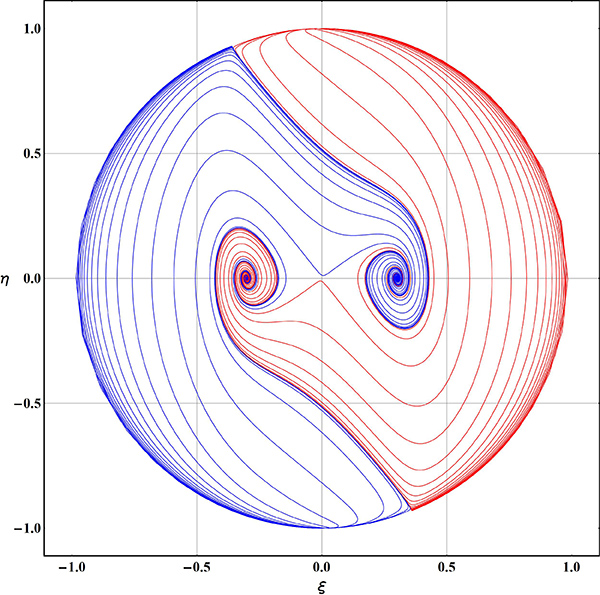}
\figcaption{\label{ris16}  The phase trajectories of the system (\ref{GrindEQ__13_}) on the Poincare sphere for the classical field with the self-action at $e_1=1$, $e_2=-1$, $\alpha_m=-10$, $\lambda_m=0.1$.}
\end{center}
\begin{center}
\includegraphics[width=8cm]{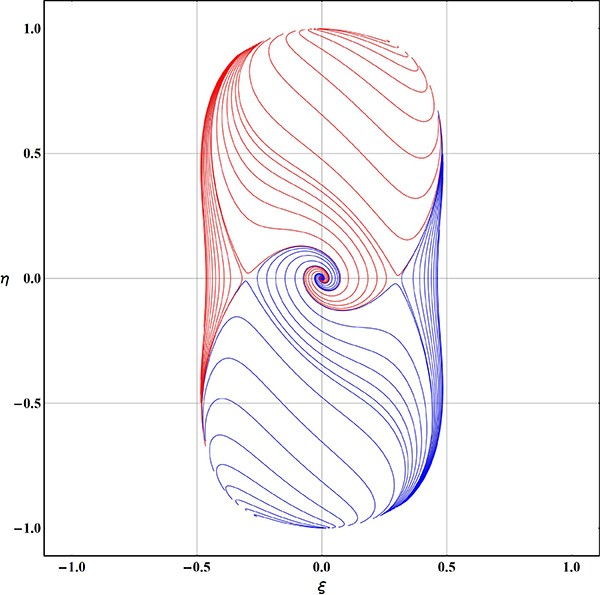}
\figcaption{\label{ris17}  The phase trajectories of the system (\ref{GrindEQ__13_}) on the Poincare sphere for the classical field with the self-action at $e_1=1$, $e_2=1$, $\alpha_m=10$, $\lambda_m=0.1$.}
\end{center}
\end{multicols}

\ruleup
\begin{center}
\includegraphics[width=16cm]{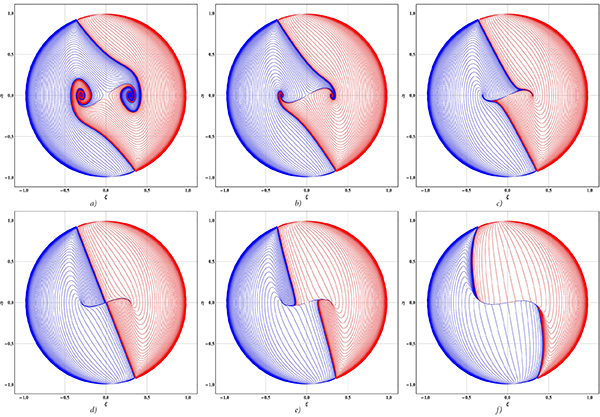}
\figcaption{\label{ris18} The phase trajectories of the system (\ref{GrindEQ__13_}) of the classical scalar field on the Poincare sphere for the parameters $e_1=1$, $e_2=-1$, $\alpha_m=-10$; the upper row, from left to right: $\lambda_m=0.1,\ 0.5,\ 1.0$; the lower row, from left to right: $\lambda_m=1.61666,\ 2.61666,\ 10$.}
\end{center}

\ruledown

\begin{multicols}{2}

\section{The Consideration of the Models with Single Scalar Fields}
First, let us notice that the coordinates on the Poincare diagrams (Fig. \ref{ris15} -- \ref{ris18}) $\xi$ and $\eta$ are normalized and connected with the dynamic variables $x$ and $y$ by the following expressions:
\[\xi=\frac{x}{\sqrt{1+x^2+y^2}};\; \eta=\frac{y}{1+\sqrt{x^2+y^2}},\]
Therefore the points on the circumference of the unitary radius correspond to infinitely great values of the dynamic variables $\Phi\to\pm\infty$, $\dot{\Phi}\to\pm\infty$ . Each certain phase trajectory is defined by a certain point on the unitary circle, and the direction of movement alongside the phase trajectory is defined by the direction from the corresponding pole to that point.

Comparing the projections of the phase portraits of the dynamic systems, based on the classical scalar field with the self-action (Fig. \ref{ris15} -- \ref{ris18}), let us note the following:

\noindent 1. The phase portraits have the central symmetry $\{\Phi,Z\}\to\{-\Phi,-Z\}$. This allows us to limit ourselves with the analysis of the phase trajectories which start in the single semiplane, for example, to be specific, in the upper one.\\
\noindent  2. In the case of the null value of the cosmological constant and absence of the self-action ($\alpha=0$) all the phase trajectories are split by the separatrices coming out from the equator  $\{\Phi_\infty=\pm\infty,Z_\infty=0\}$ into two dynamic streams coming out from the poles $\{\Phi_\infty=0,Z_\infty=\pm\infty\}$ and asymptotically winding onto the focus $\{0,0\}$ at $t\to+\infty$ (Fig. \ref{ris15}).\\
\noindent 3. In the case when $\lambda>0$, $e_2=-1$, $\alpha<0$ there appear two attractive centers and one saddle central point: the phase sepace splits into 4 dynamic streams coming out from the poles and each of the streams is split to the pair of streams; the trajectories of one of these are winding onto the left attractive center, and the second one's are winding onto the right attractive center (Fig. \ref{ris16}).\\
\noindent 4. In the case $\lambda>0$, $e_2=1$, $\alpha>0$ there appears a single attractive center and two symmetrical saddle points: the phase space splits into 6 dynamic streams where one a pair from these comes out of the poles and gets into the opposite poles while and another pair, coming out from the poles, gets back, and the third pair of streams, coming out from the poles, is winding onto the attractive center (Fig. \ref{ris17}).\\
\noindent  5. There, finally, exists one more transitional type of the phase diagram at $\lambda>0$, $e_2=-1$, $\alpha<0$   - the bifurcations when all the trajectories are split into two dynamic streams coming out from the poles and returning back after quite a complex loop. (Fig. \ref{ris18}).

Let us now consider the most interesting case of the model $\lambda\geq0$ with the classical scalar field with the self-action $e_1=1$, $e_2=-1$, $\alpha<0$, which phase trajectories are shown on Fig. \ref{ris9}. As it can be seen from the figure, the phase trajectories coming out from the infinity, approach the boundary of the area of allowable values of the dynamic variables with time. It is necessary to understand what happens with these trajectories. Let us notice that the scalar field's equation (\ref{EqField0}) for the Friedman Universe has a form of the standard equation of nonlinear oscillations with an account of the friction coefficient  $\beta(x,x')$, which is dependent on the coordinate and speed of a particle (see e.g., \cite{Ignat_16_1_stfi}):
\begin{equation}\label{Eq_dis}
x''+\beta x'+V'(x)=0,
\end{equation}
where $\beta=3H_m(x)=\sqrt{3\varepsilon_m}$. Therefore the approaching of the trajectory to the line $\varepsilon_{eff}=0$ can be interpreted as fall of the coefficient of friction to null in the neighborhood of the inaccessible range. This means that the system's trajectory should come to the trajectory of free potential nonlinear oscillations:
\begin{equation}\label{rest}
x''+V'(x)=0,
\end{equation}
which has the following integral of the total energy:
\begin{equation}\label{E0}
\frac{x'^2}{2}+V(x)=E_0.
\end{equation}
Exactly this integral describes the curve (\ref{Eq_xy4}), \emph{if we renormalize the constant $E_0$ with the help of cosmological constant}. The Fig. \ref{ris19} illustrates the results of the numerical integration of the equation of the free oscillations (\ref{rest}).
\begin{center}
\includegraphics[width=8cm]{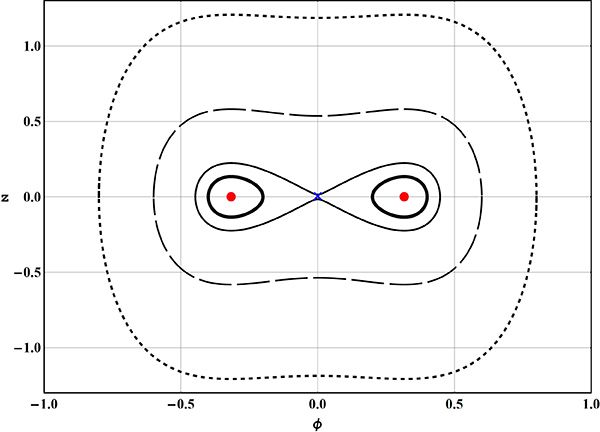}
\figcaption{\label{ris19} The phase trajectories of the free oscillations (\ref{rest}):   $e_1= 1,\ e_2=-1,\ \alpha=-10$;
 the heavy line is $\lambda = 0.288$; the thin line is $\lambda = 0.400769$; the dash-dotted line is $\lambda = 1.008$,  the dotted line is
$\lambda = 2.688$.}
\end{center}
The Fig. \ref{ris20} illustrates the results of the numerical integration of the dynamic equations (\ref{GrindEQ__13_}) in the neighborhood of the trajectory of the free oscillations.
\end{multicols}
\ruleup
\begin{center}
\includegraphics[width=16cm]{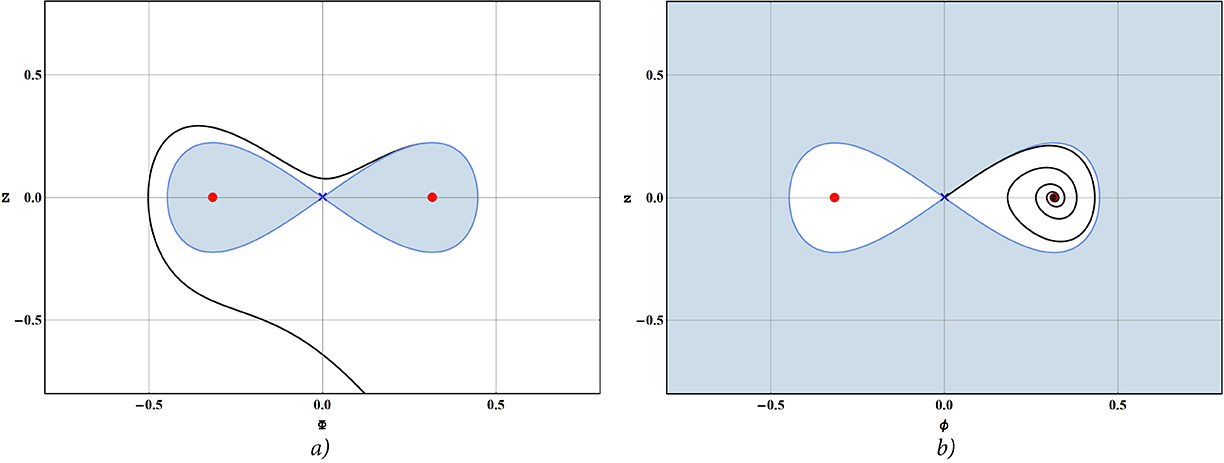}
\figcaption{\label{ris20} \textbf{a)} The approaching of the phase trajectory of the dynamic system (\ref{GrindEQ__13_}) to the trajectory of free potential oscillations (solid blue line) in the case of the classical scalar field (the solid black line): $e_1= 1,\ e_2=-1,\ \alpha=-10;\ \lambda = 0$. The permitted area is shown on the Fig. \ref{ris2}a. \textbf{b)} The dynamic system's (\ref{GrindEQ__13_}) phase trajectory's get off from the trajectory of free potential oscillations (\ref{rest}) (solid blue line) in the case of the phantom scalar field: $e_1=-1,\ e_2=1,\ \alpha=10,\lambda =0$. The permitted area is shown on the Fig. \ref{ris2}b.}
\end{center}
\ruledown
\begin{multicols}{2}
Thus, in the wide enough range of parameters of the model of nonlinear scalar field, both classical one and phantom one, the appearance of the stable boundary cycle at $t\to+\infty$ is possible; this cycle can be described by the trajectory with the null effective energy  (\ref{Eq_xy4}) and it represents the free (sustained) oscillations in the potential field. Naturally, there raises a question about the enerfy and pressure of the scalar field in this finite state. Substituting the expression for the energy density of the scalar field (\ref{varepsilon}) in the equation of the boundary trajectory (\ref{Eq_xy4}), independently on the parameters of the model we obtain for the energy density of the scalar field on the boundary curve the following expression:
\begin{equation}\label{E_l8}
\varepsilon_\infty= -\frac{\lambda}{8\pi}.
\end{equation}
Thus, this value should be negative at $\lambda>0$ which is possible \emph{only at certain values (and signs!) of the model's parameters}. Substituting the obtained value for the energy density into the expression (\ref{p}) for the scalar field's pressure, we find on the boundary curve:
\begin{equation}\label{p_l8}
p_\infty=\frac{1}{8\pi}(2m^2e_1\Phi'^2+\lambda).
\end{equation}
Let us investigate the asymptotic behavior of the system in the considered case at $\tau\to+\infty$. Since it is $H=\dot{a}/{a}\to 0$, then $\dot{a}\to o$, it follows that  $\ddot{a}\to 0$. Then, as a consequence of the Einstein equation for the space componenets it should be
$ p\to \lambda/8\pi, \; t\to+\infty$. According to (\ref{varepsilon}), (\ref{p}), (\ref{E_l8}) this is only possible when the condition $\dot{\Phi}\to0\rightarrow \Phi\to\Phi_0$ and $\varepsilon_{eff}=0$ fulfills:
\[ e_2 m^2\Phi^2_0-\frac{\alpha}{2}\Phi^4_0+\lambda=0\Rightarrow e_2 x_0^2-\frac{\alpha_m}{2}x_0^4+\lambda_m=0.\]
This equation coincides with the curve equation (\ref{Eq_xy4}) at $y=0$, consequently it defines the point lying at the intersection of the abscissa axis with this curve and can be rewritten in the terms of the potential $V(\Phi)$ (\ref{V(phi)}):
\[ V(\Phi_0)=-\frac{1}{2}\left(\frac{1}{2\alpha}+\lambda\right).\]
As it can be seen from the Fig. \ref{ris1}, there coul be two of such points $M_0(\pm \Phi_0,0)$ or four, depending on the model's parameters. Thus, such Universe can tend to the Euclidian Universe with the scalar vacuum fully compensating for the cosmological term, at $t\to+\infty$ and suitable parameters of the model. The process of the transition to this mode of expansion requires addition research.

In the case of the phantom field, apparently, we have unstable boundary cycle (see \cite{Bogoyav}), off which the phase trajectories repulse at $t\to-\infty$. For such a field the start with the Euclidian Universe and then the transition to irreversible inflation is possible. The question about the existence of the unstable boundary cycle for the phantom field also requires additional complex research.
Nevertheless, we can put forward some physical arguments in favor of the existence of the mentioned boundary cycles corresponding to null effective energy, i.e. the Euclidian Universe. For that we can analyse the behavior of the effective energy $\varepsilon_{eff}$ (\ref{E_ef}) and invariant cosmological acceleration $\Omega$ (\ref{kappa}) in the neighborhood of the trajectory of the null effective energy. Figures \ref{ris21} and \ref{ris22} illustrate the plots of these values for the model with the classical field and the figures \ref{ris23} and \ref{ris24} illustrate the plots of these values for the phantom field. These plots correspond to the phase trajectories on Fig. \ref{ris20} and are obtained on the basis of the numerical integration of the system (\ref{GrindEQ__13_}).
\begin{center}
\includegraphics[width=8cm]{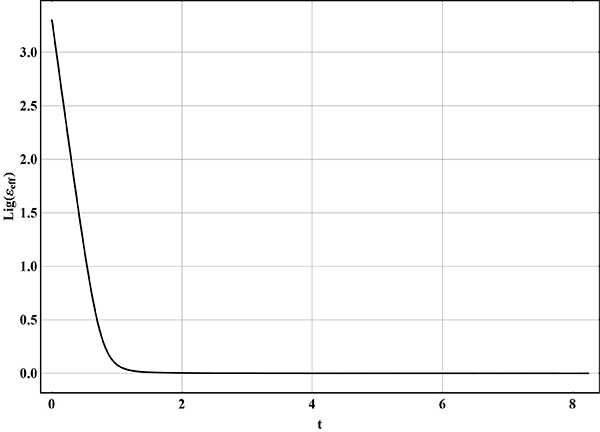}
\figcaption{\label{ris21}  The evolution of the value $8\pi\varepsilon_{eff}/m^2$, proportional to the effective energy (\ref{E_ef}) of the classical field:   $e_1= 1,\ e_2=-1,\ \alpha=-10$, $\lambda=0$, $\Phi(0)=10,\ Z(0)=0.85$. The values of the dimensionless time $\tau$ are plotted along the abscissa axis.}
\end{center}
\begin{center}
\includegraphics[width=8cm]{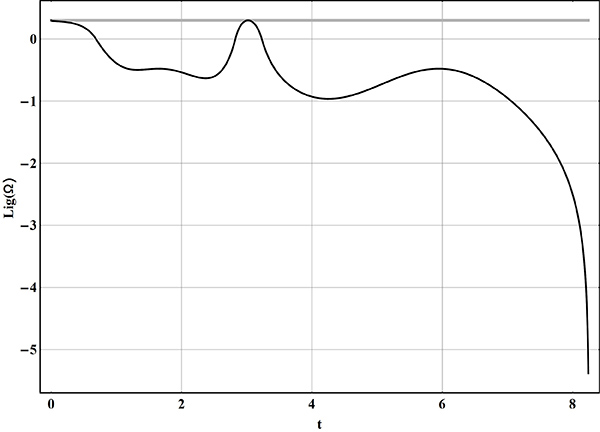}
\figcaption{\label{ris22}  The evolution of the invariant cosmological acceleration $\Omega$ (\ref{kappa}) of the classical field:   $e_1= 1,\ e_2=-1,\ \alpha=-10$, $\lambda=0$, $\Phi(0)=10, \ Z(0)=0.85$. The gray horizontal line corresponds to the value $\Omega=1$.
 }
\end{center}
For the convenience of the results representation on the entire real axis $\mathbb{R}$ having incomparable scales it was used the function $\mathrm{Lig}(x)$ \cite{Ignat_Agaf_Mih_Dima15_3_AST}:
\[\mathrm{Lig}(x)\equiv \mathrm{sgn}(x)\log_{10}(1+|x|),\]
such that:
\[\mathrm{Lig}(x) \approx \left\{%
\begin{array}{ll}%
x, &  |x|\to 0;\\
\mathrm{sgn}(x)\log_{10}|x|, & |x|\to\infty.
\end{array}
\right.
\]
\begin{center}
\includegraphics[width=8cm]{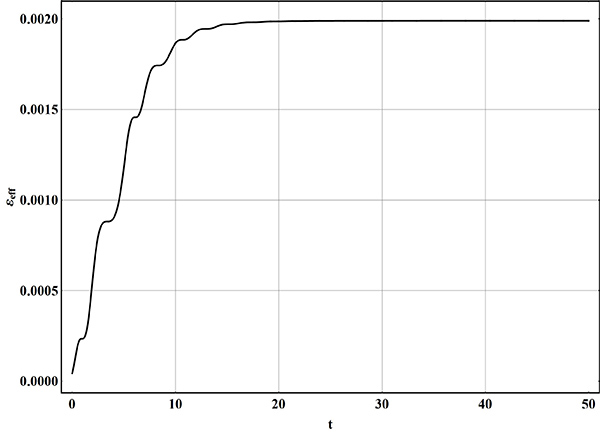}
\figcaption{\label{ris23}  The evolution of the value $8\pi\varepsilon_{eff}/m^2$, proportional to the effective energy (\ref{E_ef}) of the phantom field:   $e_1=- 1,\ e_2=1,\ \alpha=10$, $\lambda=0$, $\Phi(0)=0.3,\ Z(0)=0.22$. The values of the dimensionless time $\tau$ are plotted along the abscissa axis.}
\end{center}
\begin{center}
\includegraphics[width=8cm]{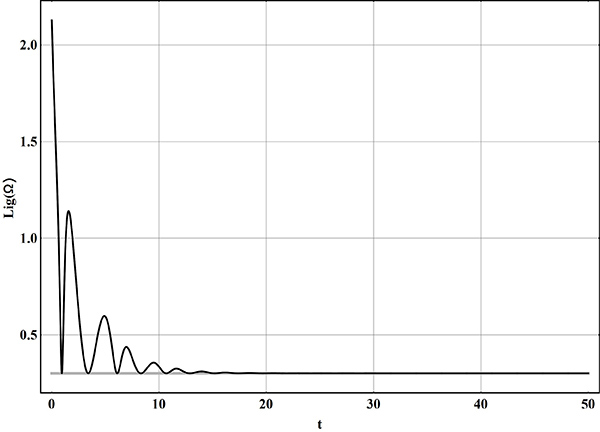}
\figcaption{\label{ris24}  The evolution of the invariant cosmological acceleration $\Omega$ (\ref{kappa}) of the phantom field:   $e_1=- 1,\ e_2=1,\ \alpha=10$, $\lambda=0$, $\Phi(0)=0.3,\ Z(0)=0.22$. The gray horizontal line corresponds to the value $\Omega=1$.
 }
\end{center}
As it can be seen on the Fig. \ref{ris21}, the value of $E=8\pi\varepsilon_{eff}/m^2$, starts from the values $E\approx 10^3$ at $\tau=0$ (to which the lower point of the phase trajectory corresponds on the Fig. \ref{ris20}.a)), and after time $\tau\sim 2$ decreases to the values close to null. Simultaneously, the value of the invariant cosmological acceleration $\Omega$, starts from the $\Omega=1$, then sharply falls at $\tau\sim 6$ and reaches great negative values $\Omega \approx -10^{5}$ at $\tau\sim 8$, which corresponds to abrupt deceleration at transition to the trajectory of free oscillations with null effective energy (see Fig. \ref{ris22}).

In the case of the phantom field (Fig. \ref{ris23}), the value of $E=8\pi\varepsilon_{eff}/m^2$, starts practically from null values at $\tau=0$ (to which the point of the phase trajectory located in the neighborhood of the saddle point $M(0,0)$ on the Fig. \ref{ris20}.b)), after time $\tau\sim 10$ reaches values $E\approx 0.002$ and then comes to the constant value. Simultaneously the magnitude of the invariant cosmological acceleration $\Omega$ starts with great positive values of $\Omega\approx 100$ at $\tau=0$ falls in the oscillation mode and at $\tau\sim 10$ tends to the constant value $\Omega =1$, which corresponds to inflation.

The unique features of the classical and the phantom scalar fields with the self-action, revealed by the Authors and concluding in the possibility of the Euclidian start of the Universe for the phantom field and the Euclidian finish of the Universe for the classical field confirm the possibility of reconsideration of the standard cosmological scenario. Apparently, this scenario can be made more elegant and the well-known contradictions can be avoided. Also, these features point to the necessity of investigation of the combined model built on the asymmetrical scalar doublet of the classical and the phantom scalar fields, which is our intent to do in the second part of the paper. Let us notice, that we have not aimed to adapt the considered models to the well-known theoretical field schemes, therefore the revealed peculiarities have, first and foremost, the fundamental nature. Let us also notice, that at linkage to certain theoretical field schemes it is necessary to take into account that all constants, as well as the time, are normalized on the mass of the scalar field $m$.


\vspace{-1mm}
\centerline{\rule{80mm}{0.1pt}}

\end{multicols}

\clearpage

\end{document}